\documentclass[sigconf,nonacm]{acmart}
\makeatletter
\global\@ACM@balancefalse
\makeatother

\setcopyright{none}
\renewcommand\footnotetextcopyrightpermission[1]{}

\usepackage{algorithmic}
\usepackage{graphicx}
\usepackage{textcomp}
\usepackage{xcolor}
\usepackage{enumitem}
\def\BibTeX{{\rm B\kern-.05em{\sc i\kern-.025em b}\kern-.08em
    T\kern-.1667em\lower.7ex\hbox{E}\kern-.125emX}}

\begin{document}

\title[UniCon: A Unified Context-Centric Modeling Paradigm for CTR Prediction]
{\texorpdfstring
  {UniCon: A Unified Context-Centric Modeling\\Paradigm for CTR Prediction}
  {UniCon: A Unified Context-Centric Modeling Paradigm for CTR Prediction}}

\author{Jiajun Cui}
\authornote{These authors contributed equally to this work.}
\affiliation{%
  \institution{Meituan}
  \city{Beijing}
  \country{China}}
\email{cuijiajun02@meituan.com}

\author{Zhengqi Xu}
\authornotemark[1]
\affiliation{%
  \institution{Meituan}
  \city{Beijing}
  \country{China}}
\email{xuzhengqi02@meituan.com}

\author{Fan Zhang}
\authornotemark[1]
\affiliation{%
  \institution{Meituan}
  \city{Beijing}
  \country{China}}
\email{zhangfan133@meituan.com}

\author{Zhangteng}
\authornote{Corresponding author.}
\affiliation{%
  \institution{Meituan}
  \city{Beijing}
  \country{China}}
\email{zhangteng09@meituan.com}

\author{Gu Tang}
\affiliation{%
  \institution{Meituan}
  \city{Beijing}
  \country{China}}
\email{gutang@sjtu.edu.cn}

\author{Honghong Zhu}
\affiliation{%
  \institution{Meituan}
  \city{Beijing}
  \country{China}}
\email{zhuhonghong@meituan.com}

\author{Mengxi Wu}
\affiliation{%
  \institution{Meituan}
  \city{Beijing}
  \country{China}}
\email{wumx0524@buaa.edu.cn}

\author{Yulin Liang}
\affiliation{%
  \institution{Meituan}
  \city{Beijing}
  \country{China}}
\email{liangyulin@meituan.com}

\author{Xingxing Wang}
\affiliation{%
  \institution{Meituan}
  \city{Beijing}
  \country{China}}
\email{wangxingxing04@meituan.com}

\begin{abstract}
Unified modeling has become a major direction for industrial
click-through rate (CTR) prediction. Existing approaches typically unify
sequential and non-sequential signals at the token level, model their
interactions in a shared backbone, and increase model capacity to improve
scaling behavior. However, this
division originates from legacy feature-engineering practice and is misaligned
with the underlying decision process. User behavior is inherently a sequence
of homogeneous context units; at the level of input organization, historical
behavior and the current request differ only in whether their outcomes are
observed or remain to be predicted. Treating them as heterogeneous signals
obscures structural dependencies within the user's decision context, limiting
both scaling efficiency and prediction quality. This limitation is particularly
pronounced in context-rich scenarios
such as e-commerce shelves and waterfall feeds. To address this, we propose
UniCon, a unified context-centric modeling architecture that treats the request
context as the basic modeling unit and organizes history and prediction targets
as homogeneous context units. Intra-context attention captures local coupling
among items within a context (\emph{Locality}), while inter-context attention
models the dynamic evolution of decision states across contexts
(\emph{Dynamics}). This organization bridges the structural gap between
history and target and supports more effective scaling of unified CTR models.
Context-unit-level sequence compression further reduces deployment overhead.
On Meituan search advertising, UniCon improves offline AUC by 0.0139
over a strong production baseline and achieves statistically significant online
lifts of 3.09\% in RPM, 2.07\% in CTR, and 2.95\% in revenue.
\end{abstract}

\ccsdesc[500]{Information systems~Recommender systems}
\ccsdesc[300]{Computing methodologies~Neural networks}

\keywords{
Click-through rate prediction, unified modeling, contextual modeling,
recommender systems, scaling laws
}

\maketitle

\section{Introduction}

The rise of large language models and foundation models has demonstrated that
the joint scaling of model capacity, data, and computation can continuously
improve model capabilities \cite{brown2020language,bommasani2021foundation,
kaplan2020scaling,hoffmann2022training}. Similar scaling trends are emerging in
industrial search, advertising, and recommendation systems, where click-through
rate (CTR) prediction remains a central task for ranking candidate items.
Recent industrial CTR models have therefore shifted from isolated expert-designed
ranking architectures toward unified modeling frameworks. These models scale
feature-interaction architectures, unify sequence modeling with feature
interaction, or build shared ranking backbones over heterogeneous industrial
inputs \cite{wukong,rankmixer,onetrans,hyformer,est}. Other industrial efforts
report related scaling gains in live ranking systems
\cite{suan,zenith,tokenmixer,mixformer}. By concentrating data and computation
in shared models, these approaches reduce redundant manual feature crosses and
task-specific modules.

Despite these advances, most unified models continue to organize inputs
according to the conventional division between sequential and non-sequential
signals. This division largely reflects legacy feature-engineering pipelines:
historical actions are encoded as a behavior sequence, whereas current
candidates, user attributes, and request signals are treated as non-sequential
ranking features, with candidates typically scored pointwise. Although
convenient for system construction, this organization is not necessarily
aligned with the underlying decision process.
Each user action occurs within a request or display context that jointly
determines the available choices and the resulting feedback. A behavior
trajectory can therefore be viewed as a sequence of structurally homogeneous
context units. Observed historical contexts and the current target context
share this internal structure; their principal distinction is whether the
outcome has already been observed or remains to be predicted. Separating them
into sequential and non-sequential branches introduces artificial
heterogeneity before model interaction begins.
This mismatch further obscures context boundaries in
strongly context-aware scenarios such as e-commerce shelves, waterfall feeds,
and search result lists. In these environments, user feedback on an item depends
not only on the item itself, but also on its local competition and
complementarity with other co-displayed items. A flat token sequence can
conflate co-occurrence within the same display list with temporal proximity
across different lists. For example, on a search results page, a user clicking
an inexpensive nearby coffee shop because the co-displayed alternatives are
more expensive or farther away does not necessarily indicate an unconditional
preference for that shop. In a later context, when a specialty coffee shop that
better matches the user's long-term preference is displayed alongside
comparable alternatives, the user may choose it instead. Preserving the two
display contexts allows the model to interpret each click relative to the
alternatives shown at that time and to track how the user's expressed
preference changes across contexts.

\begin{figure*}[t]
\centering
\includegraphics[width=0.95\textwidth]{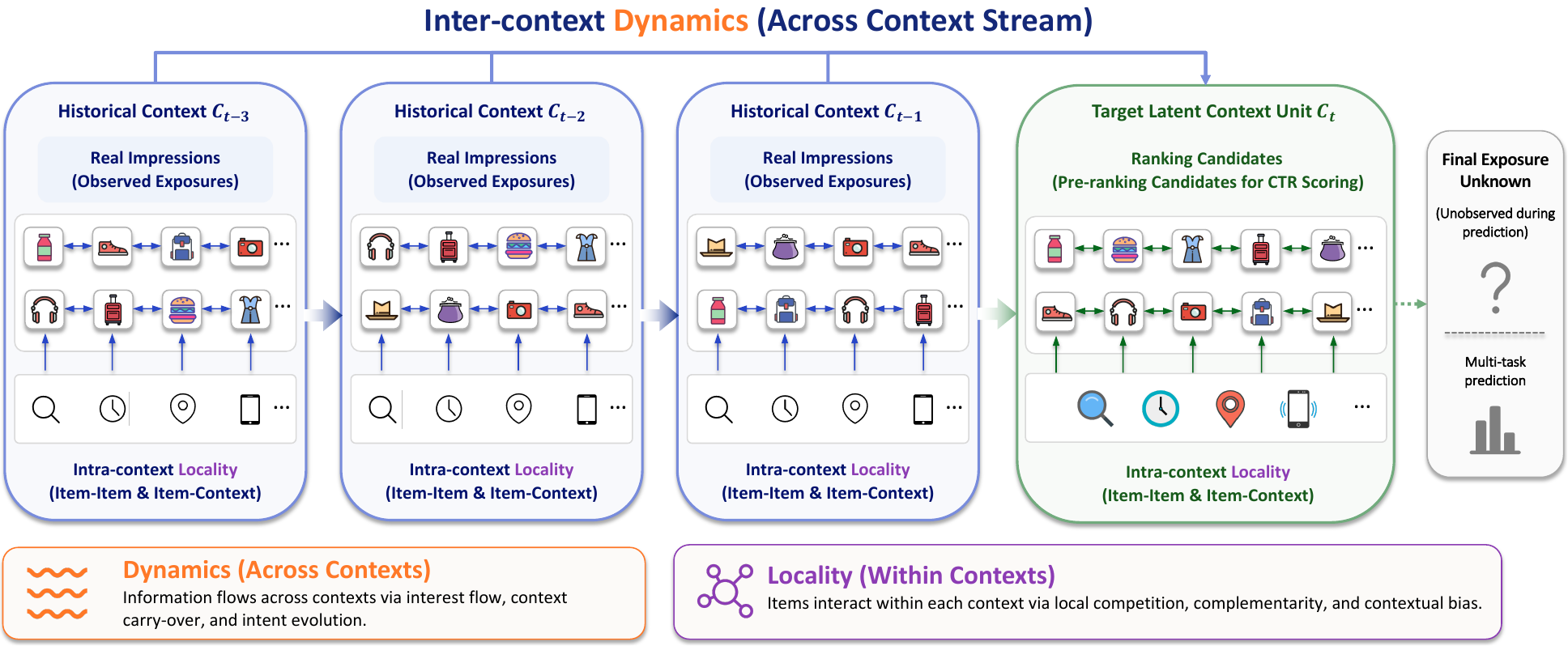}
\caption{Illustration of context locality and dynamics in CTR prediction.
Historical context units are constructed from observed display lists and
contain local interactions among items and contextual signals. The target
latent context unit is initialized from ranking candidates. Together, they form
a dynamic context sequence.}
\Description{Historical context units constructed from observed display lists
interact with the candidate-initialized target latent context unit.}
\label{fig:context}
\end{figure*}

To address these limitations, we present UniCon, a unified architecture that
extends token-level representations with context-unit organization. We define
a \emph{context unit} as the set of
items jointly displayed in one exposure event, together with the associated
user intent and environmental signals. Rather than simply inserting contextual
tokens into an existing Transformer, UniCon organizes inputs around explicit
context boundaries and builds a hierarchical interaction architecture over
these units. An \emph{intra-context} layer models interactions within each unit
to capture \emph{Locality}, including local item competition, complementarity,
and shared conditions. An \emph{inter-context} layer models interactions across
sequential units to capture \emph{Dynamics}, including the temporal evolution
of user interests and environments. At ranking time, UniCon initializes a
\emph{target latent context unit} from the current candidate set and contextual
signals because the final display list is not yet observed. During
training, exposure and absolute-position objectives refine it toward the latent
context structure reflected in the eventual display list. Historical context
units and the target unit are processed by the same hierarchical
architecture without treating the candidate set as an already observed
context. Fig.~\ref{fig:context} illustrates these local and dynamic contextual
structures.

Context units also provide natural computation boundaries for efficient
large-scale modeling. UniCon uses padding-free variable-length attention, with
context-level boundaries for intra-context interaction and input-instance-level
boundaries for inter-context interaction. It further introduces context-level
sequence compression to progressively retain historical contexts most relevant
to the target unit. Together with compiled dynamic-shape
deployment, these designs make context-centric modeling practical under
industrial training and serving constraints \cite{pytorch2}.
In Meituan search advertising, offline experiments show an AUC improvement of
0.0139 over the production base model, while an online A/B test yields
statistically significant lifts of 3.09\% RPM, 2.07\% CTR, and 2.95\% revenue.

The key contributions of this paper are summarized as follows:
\begin{itemize}[leftmargin=*]
\item \textbf{Context-centric formulation.} We define context boundaries by
grouping co-displayed items with their associated user intent
and environmental signals into context units. We then organize observed
historical contexts and current target candidates under a shared structural
schema, using a target latent context unit to bridge the observed history and
the unobserved target display context.
\item \textbf{Hierarchical contextual modeling.} We develop UniCon with
intra-context and inter-context interaction layers to capture context
\emph{Locality} and \emph{Dynamics}, respectively, unifying within-context
decision structure and cross-context evolution in a single architecture.
\item \textbf{Scalable implementation and industrial validation.} We combine
padding-free variable-length attention, context-level sequence compression,
and compiled dynamic-shape deployment to support large-scale training and
serving, and validate the resulting system through offline experiments and an
online A/B test.
\end{itemize}

\section{Related Work}

\subsection{Unified Architectures for CTR Prediction}

CTR models traditionally separate feature interaction from behavior-sequence
modeling. Wide \& Deep, DeepFM, and AutoInt model interactions among ranking
features \cite{widedeep,deepfm,autoint}, whereas DIN and DIEN extract
target-conditioned interests from behavior sequences \cite{din,dien}. Recent
unified architectures reduce this separation along two main directions.

\textbf{Scalable feature-interaction backbones.}
RankMixer, Zenith, Wukong, SUAN, and TokenMixer-Large redesign ranking
backbones for larger capacity under industrial efficiency constraints
\cite{rankmixer,zenith,wukong,suan,tokenmixer}. Their GPU-friendly operators and
scalable interaction layers support larger models. However, they focus mainly
on feature-token interaction; behavior sequences are either outside their
primary scope or handled separately, and items from the same historical
impression are not explicitly grouped.

\textbf{Unified sequential and non-sequential modeling.}
OneTrans, HyFormer, EST, and MixFormer place behavior sequences and
non-sequential features in shared or coupled Transformer-style paths
\cite{onetrans,hyformer,est,mixformer}. OneTrans uses unified tokenization with
pyramid stacking and caching, while HyFormer alternates sequence decoding and
global-feature interaction. Although these methods strengthen cross-source
interaction, their organization and acceleration remain centered on individual
behavior, feature, or query tokens, leaving impression boundaries implicit.

UniCon instead changes the unit of unification to context. It processes
observed historical lists and the target latent context unit through the same
stacked architecture, making context-level locality and cross-context
dynamics explicit while retaining a shared scalable backbone.

\subsection{Context Modeling for CTR Prediction}

Context-aware CTR models condition representations on information beyond an
isolated user--item pair. At the instance level, ContextNet uses contextual
information to refine feature embeddings, while DCIN and Deep Context Interest
Network model interactions between user behaviors and their surrounding
decision or display environments \cite{contextnet,dcin,dcin2}. Candidate-aware
methods capture another source of context: CIM encodes the candidate set into
context vectors so that CTR prediction can reflect the alternatives available
in the current request \cite{cim}. DSIN groups behavior sequences into sessions
and models interest dependencies within and across session boundaries
\cite{dsin}. Its session structure captures temporal interest organization,
whereas UniCon defines each context unit by items jointly displayed under the
same conditions.

Page- and list-aware studies further preserve structures formed during
presentation. RACP represents exposed products and page feedback as
contextualized page-wise sequences, whereas DPIN models interactions among
candidates, positions, users, and contextual signals \cite{racp,dpin}.
Classical click models likewise show that examination, position bias, and list
context affect observed clicks \cite{dbnclick,ubm,contextclick}. These studies
motivate modeling instance, candidate, page, and list context in CTR systems.

Nevertheless, most methods specialize in one context type or introduce
contextual signals while retaining behavior sequences, target items, or feature
fields as the primary input abstraction. UniCon instead treats the context
boundary as part of the model organization: it determines how signals are
grouped, how intra- and inter-context interaction is performed, and how
variable-length execution and compression are applied. In UniCon, context is
thus a structural unit of the architecture rather than an additional feature
source.

\section{Methodology}

\subsection{Overview}
\begin{figure*}[t]
\centering
\includegraphics[width=0.90\textwidth]{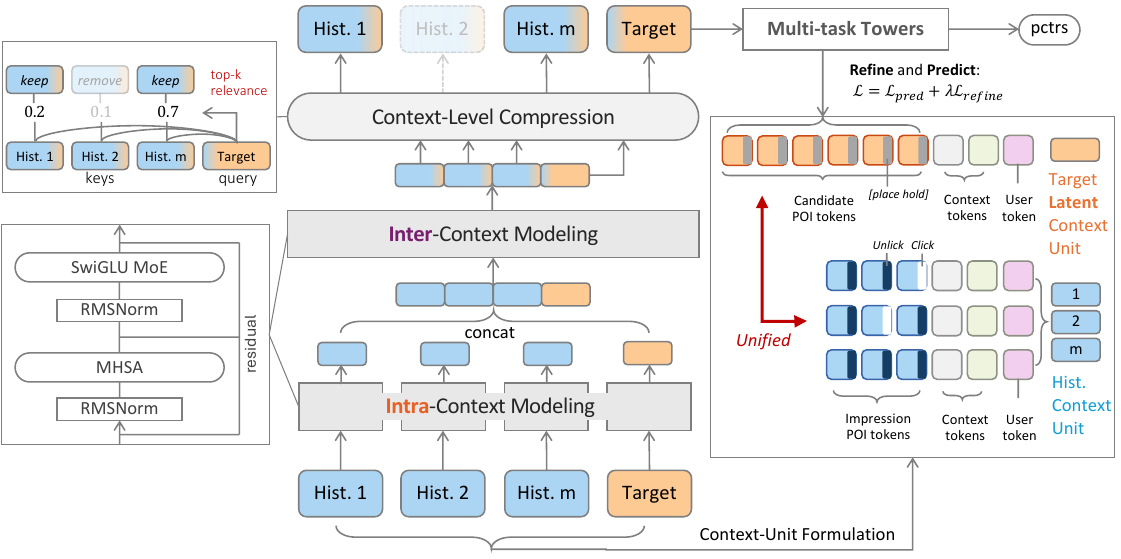}
\caption{Architecture of stacked UniConBlocks. Each context unit contains
context tokens, a user token, and item-side tokens: impression tokens for
observed historical units or candidate tokens for the target latent context
unit. Historical impression tokens include logged click feedback, whereas
candidate tokens use a learnable placeholder in the aligned feedback
component. In each UniConBlock,
the intra-context Transformer independently encodes the tokens within each
historical unit and the target unit, and the inter-context Transformer
then models their interactions. Between stacked blocks, target-aware
context compression selects the most relevant historical contexts for the next
block. As illustrated by the inset, inter-context modeling combines
self-attention with a dense MoE feed-forward sublayer. The final
representations of target candidates are fed into the CTR, exposure, and
position prediction heads.}
\Description{UniCon architecture with context-unit input organization. Each
unit contains context tokens, a user token, and either historical impression
tokens or target candidate tokens. Historical impression tokens include logged
click feedback, whereas aligned candidate tokens use learnable placeholders.
Intra-context and inter-context
Transformers are followed by context compression and multi-task prediction
heads.}
\label{fig:method-overview}
\end{figure*}

Fig.~\ref{fig:method-overview} presents the overall architecture of UniCon,
which organizes historical and target inputs as context units, alternates
intra-context and inter-context interaction in stacked UniConBlocks, and
combines target-aware context compression with multi-task prediction.

\subsection{Context-Unit Formulation}
Each input instance is organized as a sequence of observed historical context
units followed by one candidate-initialized target latent context unit. This
shared hierarchy preserves historical display structure and provides a common
schema for contextual modeling.

\subsubsection{Context-Unit Representation}
A context unit has item-side, contextual, and user-side inputs. Item-side
features include item identity, display order, and static attributes.
Contextual signals include intent, query, category preference, time, location,
device, and environment. The user-side group consists primarily of dynamic
statistical features, together with a small number of static attributes such
as gender, and is aggregated into a single user token. Because static
attributes form only a small part of this group, including them in each
context unit introduces little redundant storage.

To align the item-token schema across the history and target sides, UniCon
reserves the same feedback component in every item token. For an observed
historical unit, this component is instantiated with an embedding of the logged
click label. Because target feedback is unavailable at ranking time, the
corresponding component of each candidate token is filled with a learnable
placeholder embedding. The placeholder is optimized jointly with the
model and carries no future click information.

All features are tokenized before entering UniCon. Each item is mapped to one
item token, each type of contextual information is mapped to one context token,
and the user-side features are mapped to one user token. For a token group
\(g\), a linear tokenizer aggregates its related feature embeddings into a
unified representation:
\begin{equation}
\label{eq:tokenizer}
\begin{aligned}
\mathbf{z}_{g}
&= \mathrm{Tokenizer}_{g}\left([\mathbf{e}_{g,1}; \ldots; \mathbf{e}_{g,n_g}]\right) \\
&= \mathbf{W}_{g}[\mathbf{e}_{g,1}; \ldots; \mathbf{e}_{g,n_g}] + \mathbf{b}_{g},
\end{aligned}
\end{equation}
where \([\cdot;\cdot]\) denotes concatenation and \(\mathbf{z}_{g}\in
\mathbb{R}^{d}\). The resulting token sequence of context unit \(C\) is
\begin{equation}
\label{eq:context-token-sequence}
\mathbf{Z}_{C} =
[\mathbf{z}^{item}_{1}, \ldots, \mathbf{z}^{item}_{N_C},
\mathbf{z}^{ctx}_{1}, \ldots, \mathbf{z}^{ctx}_{K_C},
\mathbf{z}^{usr}],
\end{equation}
where \(\mathbf{z}^{usr}\) is the user token, and \(N_C\) and \(K_C\) denote
the numbers of item tokens and context tokens, respectively.

\subsubsection{Historical--Target Structural Alignment}
Under the shared schema, the history side consists of context units reconstructed
from observed exposure lists, including their item order, user-side features,
contextual signals, and feedback. The target side contains one target latent
context unit initialized from the candidates and current user-side and
contextual signals because the final display list is unavailable before
ranking. During training, multitask supervision from exposure and absolute
position further guides this candidate-initialized unit to learn the exposure
patterns and position structure of final display lists recorded in production
logs.

\subsubsection{Target-Context Supervision}
UniCon adopts an MMoE-style multi-task learning framework with multiple
prediction heads \cite{mmoe}. These objectives serve two complementary
purposes. First, because the target display list is unavailable before ranking,
exposure and absolute-position supervision refine the candidate-initialized
target latent context unit toward the latent context structure reflected in the
eventual display list. Second, by capturing which items are jointly exposed and
how they are positioned, the refined context representation provides more
informative contextual signals for CTR prediction.

The three heads predict click, exposure, and absolute position. The click head
is trained only on candidates that enter the final display list and is
conditioned on an embedding of the logged absolute position. The exposure head
is trained on the full candidate set, with a label indicating whether each
candidate enters the final exposure list. The position head is trained only on
exposed candidates and predicts their absolute exposure positions. Exposure
predictions are not fed into the click head, and click-loss gradients are
blocked from the position head. The auxiliary heads influence CTR through the
shared contextual representation. During serving, UniCon evaluates every
candidate at all feasible absolute positions to produce position-conditioned
CTR estimates.
Because exposure and position labels are generated by the incumbent production
policy, these objectives regularize the target unit toward the logged display
distribution rather than identify a policy-independent optimal context.

For each candidate \(i\) in the target unit, UniCon predicts the click
probability \(\hat{y}^{clk}_{i,p}\) at position \(p\), exposure probability
\(\hat{y}^{exp}_{i}\), and absolute-position distribution
\(\hat{\mathbf{p}}^{pos}_{i}\). The overall learning objective is:
\begin{equation}
\label{eq:learning-objective}
\mathcal{L}
= \mathcal{L}_{clk}
+ \lambda_{exp}\mathcal{L}_{exp}
+ \lambda_{pos}\mathcal{L}_{pos},
\end{equation}
where \(\mathcal{L}_{exp}\) is computed over all target candidates,
\(\mathcal{L}_{clk}\) is a binary classification loss computed over exposed
candidates at their logged absolute positions, \(\mathcal{L}_{pos}\) is the
absolute-position prediction loss computed over exposed candidates, and
\(\lambda_{exp}\),
\(\lambda_{pos}\) control the auxiliary task weights.

\subsection{Unified Contextual Modeling Architecture}

The design addresses two structural requirements: historical and target
contexts need a shared schema, while context boundaries must remain explicit as
information propagates over time. A global Transformer models cross-token
dependencies but conflates within-context co-occurrence with cross-context
proximity; independent context encoders preserve Locality but omit Dynamics.
UniCon therefore alternates intra-context and inter-context interaction over a
shared context sequence.

\subsubsection{Shared Context Sequence Representation}
After tokenization, UniCon concatenates the observed historical context units
and the candidate-initialized target latent context unit into one structured
variable-length input:
\begin{equation}
\label{eq:all-in-one-input}
\mathcal{X}
= [\mathbf{Z}_{C_1^{h}}, \mathbf{Z}_{C_2^{h}}, \ldots,
\mathbf{Z}_{C_M^{h}}, \mathbf{Z}_{C^{t}}],
\end{equation}
where each historical \(\mathbf{Z}_{C}\) is the token sequence of one observed
context unit and \(\mathbf{Z}_{C^{t}}\) is the sequence of the target latent
context unit. The model is built by stacking contextual modeling blocks:
\begin{equation}
\label{eq:stacked-unicon-blocks}
\mathbf{H}^{0} = \mathcal{X}, \qquad
\mathbf{H}^{\ell}
= \mathrm{UniConBlock}^{\ell}(\mathbf{H}^{\ell-1}), \quad
\ell=1,\ldots,L.
\end{equation}
Each block updates the complete context-unit sequence while preserving its
organization. For scoring, UniCon reads the final representation
\(\mathbf{h}^{L}_{i}\) of each candidate token in the target unit
and feeds it to three task-specific heads shared across candidates:
\begin{equation}
\label{eq:target-heads}
\hat{y}^{clk}_{i,p} = f_{clk}([\mathbf{h}^{L}_{i};\mathbf{e}^{pos}_{p}]), \quad
\hat{y}^{exp}_{i} = f_{exp}(\mathbf{h}^{L}_{i}), \quad
\hat{\mathbf{p}}^{pos}_{i} = f_{pos}(\mathbf{h}^{L}_{i}),
\end{equation}
where \(i\) denotes a target candidate,
\(\mathbf{e}^{pos}_{p}\) is the embedding of absolute position \(p\), and
\(f_{clk}\), \(f_{exp}\), and \(f_{pos}\) are the click, exposure, and
absolute-position heads, respectively. Historical and target context
units therefore pass through the same stacked architecture.
Sharing the schema does not treat the candidate pool as an observed display
list: historical units retain logged boundaries and feedback, whereas the
target unit remains latent. Instead, interaction patterns learned from observed
contexts can transfer to target representations without a separate
sequence-to-candidate fusion path.

\subsubsection{Hierarchical Intra- and Inter-Context Interaction}
Each UniConBlock uses two successive interaction levels with complementary
scopes. Intra-context interaction captures \emph{Locality} within each context
unit, whereas inter-context interaction captures \emph{Dynamics} across the
ordered context sequence.

\noindent\textbf{Intra-context interaction.}
The first level processes each historical context unit and the target latent
context unit independently. Let
\(\mathcal{C}=\{C_1^{h},\ldots,C_M^{h},C^{t}\}\) denote these units in an
input instance, and let \(\mathbf{H}_{C}^{\ell-1}\) be the token representations of
context unit \(C\) before the \(\ell\)-th block. The intra-context layer is
defined as:
\begin{equation}
\label{eq:intra-context-layer}
\widetilde{\mathbf{H}}_{C}^{\ell}
= \mathrm{IntraAttn}^{\ell}(\mathbf{H}_{C}^{\ell-1}),
\quad C \in \mathcal{C}.
\end{equation}
Attention is restricted to each context boundary, capturing Locality among its
items, feedback, user token, and contextual signals. Historical feedback is
thereby interpreted relative to the alternatives and conditions in the same
display, while target candidates are compared under the current user state and
environment. The boundary prevents adjacent displays from being treated as
local competitors.

\noindent\textbf{Inter-context interaction.}
The inter-context layer then connects the independently encoded context units.
Their intra-context outputs are concatenated in input order and passed to
inter-context attention:
\begin{equation}
\label{eq:inter-context-layer}
\mathbf{H}^{\ell}
= \mathrm{InterAttn}^{\ell}
\left([\widetilde{\mathbf{H}}_{C_1^{h}}^{\ell}, \ldots,
\widetilde{\mathbf{H}}_{C_M^{h}}^{\ell},
\widetilde{\mathbf{H}}_{C^{t}}^{\ell}]\right).
\end{equation}
The inter-context layer enables historical context units and the target unit
to exchange information about changes in user interests and display conditions.
Chronological ordering helps distinguish persistent preferences from
context-induced choices.

Both interaction layers use pre-normalized Transformer blocks with RMSNorm,
self-attention, and a SwiGLU dense MoE feed-forward layer. Intra-context
attention is restricted to each context-unit boundary, whereas inter-context
attention spans the complete input instance. The dense MoE belongs to
the contextual backbone and is distinct
from the target-side MMoE prediction heads. Appendix
\ref{app:inter-context-block} gives the complete equations for both interaction
blocks.

Alternating the two interaction levels then reinterprets cross-context
evidence within each unit, mapping Locality to intra-context interaction and
Dynamics to inter-context interaction without flattening context boundaries.

\subsection{Efficient Computation and Production Inference}
Scaling CTR models with long input histories is computationally demanding,
especially when variable-sized context units interact globally. UniCon therefore
reduces padding and unnecessary global attention while optimizing the complete
production inference path.

\subsubsection{Context-Aware Sequence Compression}
\label{sec:context-compression}
As historical exposures accumulate, the number of historical context units can
become large. If every inter-context layer attends over the full token sequence,
the complexity grows as \(O(LN^2)\), where \(L\) is the number of stacked blocks
defined in Eq.~\eqref{eq:stacked-unicon-blocks}, and \(N\) is the initial total
number of tokens in \(\mathcal{X}\) defined in Eq.~\eqref{eq:all-in-one-input}.
However, many older
historical contexts are weakly related to the current target latent context unit. UniCon
therefore first performs full inter-context interaction in the first block, and
then applies context-aware sequence compression block by block. This design
retains complete contextual modeling in the first block and progressively
shortens the sequence for subsequent global interaction.

Specifically, after block \(\ell\) (\(\ell\geq 1\)), UniCon estimates the
relevance between the target unit and each historical context.
A direct choice would be to reuse the attention scores produced by
inter-context attention. However, the variable-length attention operator follows
the FlashAttention-style online softmax computation \cite{flashattn} and does
not explicitly materialize the full attention matrix, making such scores
unavailable for selection. UniCon therefore uses a lightweight approximation: it extracts the
query vectors from the target unit and the key vectors from each historical
context, and obtains compact context-level representations by mean pooling:
\begin{equation}
\label{eq:context-compression-pool}
\mathbf{q}_{t}^{\ell}
= \mathrm{MeanPool}(\mathbf{Q}_{C^{t}}^{\ell}), \quad
\mathbf{k}_{m}^{\ell}
= \mathrm{MeanPool}(\mathbf{K}_{C_m^{h}}^{\ell}).
\end{equation}
The relevance score is then computed by query-key similarity:
\begin{equation}
\label{eq:context-compression-score}
s_m^{\ell}
= \frac{(\mathbf{q}_{t}^{\ell})^{\top}\mathbf{k}_{m}^{\ell}}{\sqrt{d}},
\quad C_m^{h} \in \mathcal{C}^{\ell-1}.
\end{equation}
According to these scores, UniCon keeps a fixed ratio \(r_{\ell}\) of
historical context units at block \(\ell\), where the number of retained units
is \(K_{\ell}=\lceil r_{\ell}|\mathcal{C}^{\ell-1}_{h}|\rceil\):
\begin{equation}
\label{eq:context-compression-select}
\begin{array}{@{}l@{\;}c@{\;}l@{}}
K_{\ell}
&=& \left\lceil r_{\ell}|\mathcal{C}^{\ell-1}_{h}| \right\rceil, \\
\mathcal{S}^{\ell}
&=& \mathrm{TopK}(\{s_m^{\ell}\mid C_m^{h}\in\mathcal{C}^{\ell-1}_{h}\}, K_{\ell}), \\
\mathcal{C}^{\ell}_{keep}
&=& \{C_m^{h} \mid m \in \mathcal{S}^{\ell}\} \cup \{C^{t}\}.
\end{array}
\end{equation}
During training, UniCon uses Gumbel-TopK with a straight-through estimator. The
forward pass applies a hard TopK mask and prunes unselected context units,
whereas the backward pass uses gradients from the continuous relaxation.
Serving applies the same hard TopK selection without the relaxation. The
retained set becomes the input context set of the next block, i.e.,
\(\mathcal{C}^{\ell}=\mathcal{C}^{\ell}_{keep}\). Let \(N_{\ell}\) denote the
number of tokens retained after the \(\ell\)-th compression step, with
\(N_0=N\). The same retention ratio \(r_{\ell}\) approximately controls token
length, i.e., \(N_{\ell}\approx r_{\ell}N_{\ell-1}\). Since the first block is
computed on the full sequence, the dominant global attention cost is reduced
from:
\begin{equation}
\label{eq:compression-complexity}
O(LN^2) \quad \text{to} \quad
O\left(N^2\sum_{\ell=0}^{L-1}\prod_{j=1}^{\ell} r_j^2\right),
\end{equation}
where the product term is defined as \(1\) when \(\ell=0\), corresponding to
the full first block.
When a fixed retention ratio is used for simplicity, i.e.,
\(r_1=r_2=\cdots=r_{L-1}=r\), this becomes
\(O(N^2\sum_{\ell=0}^{L-1}r^{2\ell})\). This geometric series is bounded
by \(1/(1-r^2)\), yielding an attention-cost upper bound of
\(O(N^2/(1-r^2))\) instead of \(O(LN^2)\).

The serving implementation fuses score computation, hard TopK selection, and
historical-context gathering instead of materializing their intermediate
results through separate operators.
The fused operator directly produces the compact context sequence consumed by
the next UniConBlock, reducing intermediate memory traffic and kernel-launch
overhead. These savings recur between successive global-interaction blocks.

\subsubsection{Variable-Length Attention Operator}
\label{sec:varlen-attention}
In a standard Transformer implementation, variable context lengths are usually
handled by padding all contexts to a fixed length and constructing customized
attention masks. This design wastes computation on invalid tokens and enforces
context boundaries only through the mask. UniCon
instead implements both intra-context and inter-context layers with a
variable-length attention operator, where tokens are stored as contiguous
segments and attention boundaries are specified by segment offsets. The operator
follows the IO-aware exact attention principle of FlashAttention \cite{flashattn}
while exposing segment offsets to represent context boundaries.

The segment definition changes with the interaction stage. For intra-context
modeling, each context unit is one variable-length segment:
\begin{equation}
\label{eq:varlen-intra}
\mathcal{B}_{intra}
= \{ \mathrm{span}(C_1^{h}), \ldots, \mathrm{span}(C_M^{h}),
\mathrm{span}(C^{t}) \}.
\end{equation}
Attention is computed independently inside each segment, so tokens from
different context units cannot attend to each other in the intra-context layer
without relying on dense custom masks. For inter-context modeling, the segment
boundary is switched to the input-instance level:
\begin{equation}
\label{eq:varlen-inter}
\mathcal{B}_{inter}
= \{ \mathrm{span}(C_1^{h}, \ldots, C_M^{h}, C^{t}) \}.
\end{equation}
All tokens in the same input instance can then interact in the inter-context
layer. The two segment definitions implement context-unit isolation for
Locality and instance-level interaction for Dynamics without padding either
stage.

\subsubsection{Production Inference}
\label{sec:production-inference}
Beyond the context-aware operators, we optimize the complete serving path at
the feature, operator, and runtime levels.

\textbf{Candidate sharding.}
The production implementation limits each target-side shard to 300 candidates.
A pool of \(N_{\mathrm{cand}}\) candidates is hash-partitioned into
\(\lceil N_{\mathrm{cand}}/300 \rceil\) shards, each independently forming a
target latent context unit; their position-conditioned scores are merged after
inference. Training and serving use the same procedure.
Appendix~\ref{app:sharding-analysis} analyzes the resulting
effectiveness--efficiency trade-off.

\textbf{Feature extraction.}
We design a \emph{FeatureColumnCollector} that consolidates previously
scattered feature-extraction logic into a unified, graph-compatible component.
It constructs model inputs from registered feature columns and removes
non-serializable Python-side extraction paths that otherwise hinder graph
export. Centralizing these operations also avoids redundant feature collection
and transformation, improving the efficiency of the feature-extraction stage.

\textbf{Parallel dense MMoE execution.}
The multitask prediction heads use a dense MMoE structure, for which a naive
implementation launches expert MLPs separately. We parallelize expert
computation with cuBLAS primitives and custom CUDA kernels. For the first
projection, the weights of all experts are packed and evaluated through a fused
matrix multiplication. Because the intermediate SwiGLU activation prevents the
two MLP projections from being collapsed into a single linear operation, the
second projection uses batched matrix multiplication to process multiple
experts in parallel. This implementation preserves the dense MMoE semantics
while reducing operator-dispatch and intermediate-memory overhead.

\textbf{Ahead-of-time compiled serving.}
We export the complete inference graph with AOTInductor and execute the compiled
artifact through a C++ runtime, removing Python from the online inference path
\cite{pytorch2}. Dynamic-shape compilation preserves variable-length execution
for requests with different numbers of contexts, candidates, and context
tokens without reverting to fixed padded inputs.

\section{Experiments}
\subsection{Dataset and Metrics}
Our experiments are conducted on Meituan search advertising data collected from
a one-year chronological production window. The two immediately following days
are used as the validation and test sets, respectively. The three splits have
no overlapping requests, and all features are constructed without future
information. Both the number of unique users and the number of unique items are
on the order of hundreds of millions. Each input history can contain hundreds
of context units, with fewer than ten items per unit, while each production
candidate shard contains up to 300 candidates.

For offline evaluation, we use AUC to measure global ranking quality and
LogLoss to measure probabilistic calibration. GAUC is computed at the request
level over exposed candidates. Requests containing only positive or only
negative labels are excluded because their AUC is undefined, and the remaining
request-level AUC values are weighted by their sample counts. For online
evaluation, we report RPM, defined as revenue per thousand impressions, CTR to
measure click-through effectiveness, and total revenue to capture overall
business impact.

\subsection{Baselines}
We organize the baselines into three groups.
\textbf{Production baseline.} Base is the online production ranking model. It combines DIN-based
behavior modeling \cite{din}, SENET-style feature recalibration
\cite{fibinet}, and DCN-V2 feature crossing \cite{dcnv2} within a composite
ranking architecture.
\textbf{Scaling baselines.} OneTrans and HyFormer are unified CTR
architectures that jointly model sequential behaviors and non-sequential
ranking features \cite{onetrans,hyformer}, while RankMixer scales ranking
through hardware-efficient feature mixing \cite{rankmixer}.
\textbf{Scaling + context baselines.} We further strengthen these backbones
with established context modules, yielding \textbf{OneTrans+CIM},
\textbf{HyFormer+DSIN+CIM}, and \textbf{RankMixer+DSIN+CIM}. DSIN provides
session-aware historical modeling, whereas CIM introduces candidate-aware
contextual interaction \cite{dsin,cim}. We do not combine DSIN with OneTrans
because DSIN would duplicate or replace OneTrans's native historical-sequence
path rather than act as an orthogonal context module; CIM augments
candidate-side context without changing that path. All baselines use the same
data, raw feature fields, and training resources as UniCon. Architectures that
do not natively support context hierarchy receive the same fields organized
according to their original designs. We sweep multiple parameter settings for
each research baseline and report the highest-AUC configuration in
Table~\ref{tab:main-offline}.

\subsection{Experimental Setup}
All models are trained on 16 NVIDIA A100 GPUs with a batch size of 1600 and
AdamW \cite{adamw}. The token dimension is 512 with four attention heads;
UniCon-Small, UniCon-Mid, and UniCon-Large stack 3, 6, and 12 UniConBlocks.
The scaling study uses uncompressed models, while the main offline and online
experiments use compressed UniCon-Large with \(r_{\ell}=0.5\), using
Gumbel-TopK with straight-through estimation and forward hard pruning during
training, and hard TopK during serving. GFLOPs are measured at the average input
length, and all baselines are selected on the same validation set. AUC
differences within 0.0003 are treated as normal run-to-run variation, while an
AUC improvement greater than 0.0003 is regarded as a clear offline gain.

The complete candidate pool entering fine ranking is retained without negative
sampling. Candidate order within each shard is randomized, and neither scores
nor ranking positions from a previous model are used as input. Training and
serving use the same hash-based sharding procedure. Historical context units
are ordered chronologically; when the configured maximum length is exceeded,
the most recent units are retained before context-aware compression.

\subsection{Overall Performance}
Table~\ref{tab:main-offline} compares prediction quality, parameter count, and
inference GFLOPs.

\begin{table}[t]
\caption{Overall offline performance comparison.}
\label{tab:main-offline}
\centering
\setlength{\tabcolsep}{2.4pt}
\begin{tabular}{lccc|c}
\hline
Model & AUC & GAUC & LogLoss & Params/FLOPs \\
\hline
Base & 0.8558 & 0.8076 & 0.2084 & 0.09B/9.69G \\
\hline
OneTrans & 0.8647 & 0.8158 & 0.2021 & 0.21B/195.61G \\
HyFormer & 0.8627 & 0.8140 & 0.2035 & 0.15B/130.76G \\
RankMixer & 0.8646 & 0.8157 & 0.2024 & 0.40B/156.25G \\
\hline
OneTrans+CIM & 0.8657 & 0.8162 & 0.2017 & 0.21B/208.75G \\
HyFormer+DSIN+CIM & 0.8653 & 0.8160 & 0.2015 & 0.16B/145.97G \\
RankMixer+DSIN+CIM & 0.8661 & 0.8171 & 0.2013 & 0.40B/171.46G \\
\hline
UniCon-Small & 0.8683 & 0.8184 & 0.2001 & 0.09B/201.60G \\
UniCon-Mid & 0.8693 & 0.8190 & 0.1995 & 0.17B/401.35G \\
UniCon-Large\textsuperscript{\dag} & 0.8697 & 0.8194 & 0.1991 & 0.33B/197.14G \\
\hline
\end{tabular}
\par\vspace{2pt}
\raggedright\footnotesize
\textsuperscript{\dag}\,Denotes the context-compressed version used for production
serving; compression is enabled during both training and inference, reducing
computation from 801.29 to 197.14 GFLOPs.
\vspace{-0.5em}
\end{table}

UniCon outperforms both the scaling baselines and their context-enhanced
variants on all three offline metrics. Adding DSIN or CIM improves the
corresponding backbones, but these combinations still organize sequence and
target-side context through separate or attached modules. In contrast,
UniCon-Small already exceeds every research baseline in AUC, showing that the
context-unit organization and hierarchical interaction are more effective than
attaching context modules to existing backbones. Performance further improves
with model capacity, and the compressed production UniCon-Large retains the
strongest overall quality at 197.14 GFLOPs.

\subsection{Ablation Studies}
We conduct ablation studies on the compressed UniCon-Large configuration to
isolate the contribution of each core component. The first ablation removes
explicit context organization while preserving contextual information: each
context-token representation is added to its associated item tokens, and the
resulting item tokens are processed as a flattened sequence. This variant uses
the same parameter count and comparable GFLOPs as UniCon, thereby isolating the
effect of explicit context boundaries. The second ablation retains historical
context units but decomposes the target-side context representation into
separate tokens for individual non-sequential features, breaking the shared structural
organization between the historical and target sides. The third ablation
removes hierarchical context modeling and instead processes the flattened
historical and target tokens with a parameter-matched global token-level
backbone. The fourth
removes only intra-context interaction. Each removed intra-context layer is
replaced with an additional inter-context layer, preserving both the total
Transformer depth and parameter count. We further ablate the exposure and
absolute-position objectives separately and jointly, and compare context
compression with no compression and time-based truncation.

\begin{table}[t]
\caption{Ablation study of UniCon-Large.}
\label{tab:ablation}
\centering
\begin{tabular}{lccc}
\hline
Variant & AUC & GAUC & LogLoss \\
\hline
UniCon-Large & 0.8697 & 0.8194 & 0.1991 \\
w/o Context Organization & 0.8687 & 0.8179 & 0.1997 \\
w/o Target Context Unification & 0.8690 & 0.8189 & 0.1998 \\
w/o Context Modeling & 0.8637 & 0.8150 & 0.2032 \\
w/o Intra-Context & 0.8673 & 0.8168 & 0.2004 \\
w/o Exposure Loss & 0.8665 & 0.8172 & 0.2012 \\
w/o Position Loss & 0.8693 & 0.8189 & 0.1996 \\
w/o Aux. Loss & 0.8663 & 0.8170 & 0.2013 \\
w/o Compression & 0.8698 & 0.8195 & 0.1991 \\
Time Truncation & 0.8691 & 0.8191 & 0.1994 \\
\hline
\end{tabular}
\vspace{-0.5em}
\end{table}

The ablations isolate UniCon's main modeling choices. Removing context
organization worsens all three metrics, showing the value of explicit context
membership. Decomposing the target-side context representation into separate
feature tokens also lowers performance, indicating that a shared context-unit interface
reduces the structural mismatch between historical and target inputs. Flattening
historical and target tokens in a parameter-matched backbone without
hierarchical context modeling causes the largest degradation. Removing only
intra-context interaction also degrades all metrics despite matched depth and
parameter count, isolating the benefit of Locality modeling from additional
capacity.

Exposure prediction contributes more than absolute-position prediction, though
both improve the target-side representation; removing both produces the
largest drop among the auxiliary-task ablations. For long-context handling,
context compression matches the uncompressed model within normal run-to-run
variation while using far less computation. Its advantage over time-based
truncation also favors target-aware selection over a fixed recency rule.

\subsection{Scaling Experiments}
\begin{figure}[t]
\centering
\includegraphics[width=0.78\columnwidth]{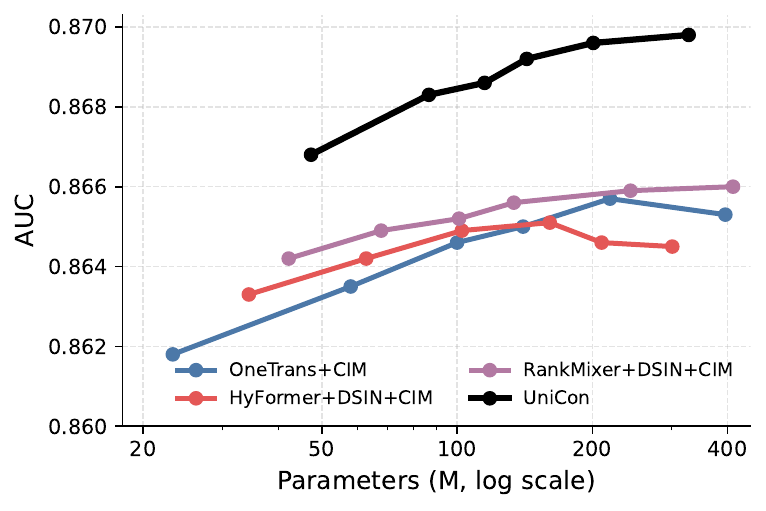}
\caption{Model scaling comparison across parameter scales. The
x-axis reports parameter count in millions on a logarithmic scale, and the y-axis reports AUC.}
\Description{AUC comparison across parameter scales for UniCon and representative
baseline models.}
\label{fig:param-scaling}
\vspace{-1.5em}
\end{figure}

Fig.~\ref{fig:param-scaling} compares uncompressed UniCon models with
representative unified and context-enhanced scaling baselines.

UniCon shows stable gains as parameter size grows
from approximately 0.05B to 0.33B. AUC increases from 0.8668 to 0.8698. At the same
0.33B parameter count, context compression reduces computation from 801.29
to 197.14 GFLOPs while retaining an AUC of 0.8697. The scaling curve in
Fig.~\ref{fig:param-scaling} further shows that UniCon maintains a
clear advantage over the representative unified and scaling-oriented baselines
across comparable parameter ranges. The consistent gap suggests that
context-unit organization remains useful as model capacity increases rather
than benefiting only one operating point.

\subsection{Compression Trade-off}
\label{sec:compression-tradeoff}
We vary the fixed context keep ratio \(r\) for UniCon-Large under the same
evaluation setting and average input length.
\begin{figure}[t]
\centering
\includegraphics[width=0.95\columnwidth]{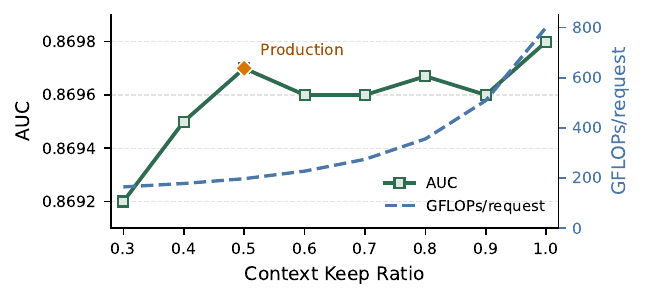}
\caption{Context keep ratio versus UniCon-Large AUC and measured GFLOPs.}
\Description{UniCon-Large AUC remains stable at moderate context keep ratios,
while measured GFLOPs per request decrease rapidly as the ratio is reduced.}
\label{fig:compression-ratio}
\vspace{-1.5em}
\end{figure}
AUC remains between 0.8696 and 0.8698 for \(r=0.5\) to \(1.0\). Reducing
\(r\) from 1.0 to the production setting of 0.5 changes AUC from 0.8698 to
0.8697 while reducing profiled computation by 75.4\%. More aggressive
compression leaves the final blocks with nearly a single historical context unit,
weakening cross-context interaction; AUC correspondingly declines to 0.8695 at
\(r=0.4\) and 0.8692 at \(r=0.3\).

\subsection{Serving Efficiency}
On production UniCon-Large, we incrementally enable Variable-Length Attention
(Sec.~\ref{sec:varlen-attention}) and Context-Aware Sequence Compression
(Sec.~\ref{sec:context-compression}) from a fixed-length, uncompressed baseline.
Variable-length attention improves throughput by 64.5\% without changing
offline AUC; compression raises the gain to 258.1\% over the baseline, or
117.6\% over variable-length attention alone, with a 0.0001 AUC difference.
Despite higher GFLOPs than the production Base, the deployment meets the
latency SLA.

\subsection{Online A/B Testing}
A seven-day A/B test in Meituan search advertising assigns 20\% of traffic to
UniCon-Large, and reports cumulative changes over the full experiment. RPM
measures revenue per thousand impressions, whereas revenue measures total
income, separating monetization efficiency from aggregate business impact.
UniCon-Large achieves lifts of 3.09\% RPM, 2.07\% CTR, and 2.95\% revenue.
All three are significant under two-sided tests (\(p\leq 0.01\)). Latency
increases relative to the Base but remains within the production SLA, with no
significant degradation in other guardrail metrics.

\section{Conclusion}
UniCon organizes observed history and current candidates under a shared
context-unit schema. Intra-context and inter-context layers capture Locality and
Dynamics, while auxiliary supervision refines the target representation.
UniCon outperforms unified and context-enhanced baselines across parameter
scales; variable-length attention, context compression, and compiled serving
keep it within the production SLA. Offline and online gains support
context-level unification for context-rich industrial ranking.

\appendix
\section{Intra- and Inter-Context Block Implementation}
\label{app:inter-context-block}

Both interaction stages use the pre-normalized residual block shown in
Fig.~\ref{fig:method-overview}. For stage
\(s\in\{\mathrm{intra},\mathrm{inter}\}\), let \(\mathbf{X}_{s}^{\ell}\) be
its input. The attention and feed-forward updates are
\begin{equation}
\label{eq:appendix-context-block}
\begin{aligned}
\mathbf{U}_{s}^{\ell}
&= \mathbf{X}_{s}^{\ell}
 + \mathrm{MHSA}_{s}^{\ell}\!\left(
 \mathrm{RMSNorm}_{s,1}^{\ell}(\mathbf{X}_{s}^{\ell})\right),\\
\mathcal{B}_{s}^{\ell}(\mathbf{X}_{s}^{\ell})
&= \mathbf{U}_{s}^{\ell}
 + \mathrm{DenseMoE}_{s}^{\ell}\!\left(
 \mathrm{RMSNorm}_{s,2}^{\ell}(\mathbf{U}_{s}^{\ell})\right).
\end{aligned}
\end{equation}
Thus, each sublayer is pre-normalized and followed by a residual connection
\cite{rmsnorm}. The two stages use the same block structure but different
attention boundaries.

For intra-context modeling, Eq.~\ref{eq:appendix-context-block} is applied
independently to every context unit:
\begin{equation}
\label{eq:appendix-intra-block}
\widetilde{\mathbf{H}}_{C}^{\ell}
= \mathcal{B}_{\mathrm{intra}}^{\ell}
  (\mathbf{H}_{C}^{\ell-1}),
\qquad C\in\{C_1^h,\ldots,C_M^h,C^t\}.
\end{equation}
The same intra-context parameters are shared across these units, while segment
offsets prevent attention from crossing their boundaries.

For inter-context modeling, the locally encoded units are first concatenated:
\begin{equation}
\label{eq:appendix-inter-input}
\mathbf{G}^{\ell}
= [\widetilde{\mathbf{H}}_{C_1^h}^{\ell},\ldots,
   \widetilde{\mathbf{H}}_{C_M^h}^{\ell},
   \widetilde{\mathbf{H}}_{C^t}^{\ell}],
\qquad
\mathbf{H}^{\ell}
= \mathcal{B}_{\mathrm{inter}}^{\ell}(\mathbf{G}^{\ell}).
\end{equation}
Here the attention boundary is the complete input instance, enabling
cross-context communication before expert transformation.

Each dense MoE contains \(E\) SwiGLU experts \cite{swiglu}. Omitting biases for
clarity, expert \(e\) and the dense mixture are
\begin{equation}
\label{eq:appendix-dense-moe}
\begin{aligned}
\mathcal{E}_{e}(\mathbf{X})
&= \left[
 \mathrm{SiLU}(\mathbf{X}\mathbf{W}_{g,e})
 \odot (\mathbf{X}\mathbf{W}_{u,e})
 \right]\mathbf{W}_{d,e},\\
\boldsymbol{\Pi}(\mathbf{X})
&= \mathrm{softmax}(\mathbf{X}\mathbf{W}_{r}),\\
\mathrm{DenseMoE}(\mathbf{X})
&= \sum_{e=1}^{E}
 \boldsymbol{\Pi}_{:,e}(\mathbf{X})\odot\mathcal{E}_{e}(\mathbf{X}).
\end{aligned}
\end{equation}
Unlike sparse expert routing, all experts participate in the dense mixture.
The mixture weights and expert transformations are learned jointly with the
contextual backbone.

\section{Impact of Candidate Sharding}
\label{app:sharding-analysis}

Full-set modeling preserves all target-side interactions but scales in
computation and memory with the candidate count. Sharding bounds this cost and
enables independent execution while restricting interactions to each shard. We
vary only the maximum shard size for UniCon-Large, keeping the architecture and
evaluation setting fixed.

\begin{figure}[t]
\centering
\includegraphics[width=0.95\columnwidth]{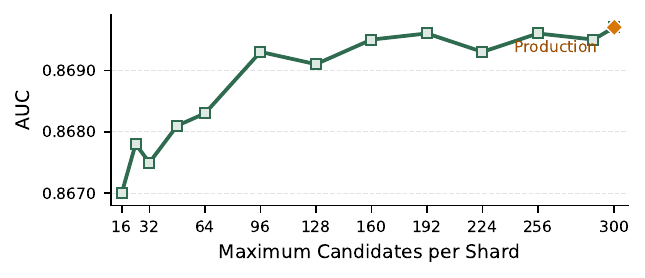}
\caption{Candidate shard size versus UniCon-Large AUC.}
\Description{AUC remains close to 0.8697 for large candidate shards and
generally decreases as the maximum shard size becomes small.}
\label{fig:shard-analysis}
\end{figure}

As shown in Fig.~\ref{fig:shard-analysis}, a shard size of 300 achieves
0.8697 AUC, and all settings from 160 to 300 remain within 0.0004 of this
result. AUC falls to 0.8683 at 64 candidates and 0.8670 at 16 candidates:
smaller shards more often separate related or competing candidates, weakening
the contextual evidence available to the target latent context unit.

The production maximum is therefore 300; larger requests are hash-partitioned
under the same training and serving rule.

\end{document}